\documentclass{emulateapj}

\usepackage{hyperref}
\usepackage{amsmath}

\shorttitle{}
\shortauthors{Jaeggli \& Norton}

\newcommand{\alp}{$\alpha$}
\newcommand{\bet}{$\beta$}
\newcommand{\gam}{$\gamma$}
\newcommand{\delt}{$\delta$}

\begin{document}

\title{The Magnetic Classification of Solar Active Regions 1992 - 2015}

\author{S.~A. Jaeggli\altaffilmark{1}}
\affil{NASA Goddard Space Flight Center, Solar Physics Laboratory, Code 671, Greenbelt, MD 20771, USA}

\author{A.~A. Norton}
\affil{W.~W. Hansen Experimental Physics Laboratory, Stanford University, Palo Alto, CA 94305, USA}

\email{sarah.jaeggli@nasa.gov}

\altaffiltext{1}{NASA Postdoctoral Program Fellow}

\begin{abstract}
The purpose of this letter is to address a blind-spot in our knowledge of solar active region statistics.  To the best of our knowledge there are no published results showing the variation of the Mount Wilson magnetic classifications as a function of solar cycle based on modern observations.  We show statistics for all active regions reported in the daily Solar Region Summary from 1992 January 1 to 2015 December 31.  We find that the \alp\ and \bet\ class active regions (including all sub-groups e.g. \bet\gam, \bet\delt) make up fractions of approximately 20\% and 80\% of the sample respectively.  This fraction is relatively constant during high levels of activity, however, an increase in the \alp\ fraction to about 35\% and and a decrease in the \bet\ fraction to about 65\% can be seen near each solar minimum and is statistically significant at the 2-$\sigma$ level.  Over 30\% of all active regions observed during the years of solar maxima were appended with the classifications \gam\ and/or \delt, while these classifications account for only a fraction of a percent during the years near the solar minima.  This variation in the active region types indicates that the formation of complex active regions may be due to the pileup of frequent emergence of magnetic flux during solar maximum, rather than the emergence of complex, monolithic flux structures.
\end{abstract}

\keywords{Sun: activity---magnetic field---sunspots}

\section{Introduction}
Magnetic classifications provide a simple way to describe the configuration of the magnetic flux and sunspots in a solar active region (AR).  The Mount Wilson (or Hale) classification system for sunspot groups put forward by \citet{hale19} has been used for nearly a century.  In the original Hale classification scheme, the designation ($\boldsymbol\alpha$) is given to regions that contains a single sunspot or sunspot group all having the same polarity.  Generally these also have a weaker opposite polarity counterpart which is not strong or concentrated enough to produce sunspots.  ($\boldsymbol\beta$) is assigned to regions which have two sunspots or sunspot groups of opposite polarity.  The classification ($\boldsymbol\gamma$) is appended to the above classes to indicate the AR has a complex region of sunspots with intermixed polarity.  This classification can also be used individually to describe an AR that has no organized magnetic behavior.  As an addendum to the original scheme, \citet{kunzel65} proposed an additional classification to modify the existing three.  ($\boldsymbol\delta$) indicates that at least one sunspot in the region contains opposite magnetic polarities inside of a common penumbra separated by no more than $2^\circ$ in heliographic distance (24 Mm or $33''$ at disk center).

Although the Mount Wilson magnetic configuration provides a limited description of the complexity of ARs, the categories are sufficiently general to encompass nearly every AR that occurs.  As a classification scheme based on human perception of symmetry, it is highly subjective, but it is also more easily determined than other measures of magnetic complexity such as fractal dimensions \citep{mcateer05}.

\citet{hale38} presented the statistics of 2174 sunspot group classifications that were assigned based on visible light imaging and spectropolarimetric measurements taken at Mt. Wilson with the 60-ft and 150-ft solar towers almost every day between 1915 and 1924, covering the majority of solar cycle 15.  On average for each year, 38\% were \alp, 58\% were \bet, 4\% were \bet\gam, 1\% were \gam, and 7\% were unclassified.  No \delt\ classifications were given as the study of \citet{hale38} was conducted before the introduction of the \delt\ class.  The classifications show only a few percent variation as a function of year and there is no significant behavior as a function of solar cycle.

It has been known for a half a century that magnetic configurations that contain a \delt\ have predispositions to producing flares \citep{kunzel60}.  \citet{smith68} used a scheme similar to Hale's to classify magnetic regions associated with sunspots in Mount Wilson magnetogram data observed on a daily basis between 1959 and 1962, near the peak of solar cycle 19.  They found that the more complex sunspot classes correlate with increased flare activity.  Since then there have been numerous statistical and case studies that investigate the properties of ARs which host flaring activity. \citet{atac87}, \citet{sammis00}, \citet{ternullo06}, and \citet{guo14} all confirm statistically that classes that contain \delt-spots are the most flare productive, but flare productivity has been found to be correlated with other properties such as sunspot area, total magnetic flux, region lifetime, and various other measures of AR and sunspot complexity.  However, these studies neglect reporting the base distribution from which complex, flare productive regions arise.

Following \citet{hale38},  to the best of our knowledge, there has been no simple, statistical study of AR classes based on modern observations taken over a full solar cycle.  However, due to modern synoptic observations and the availability of reported information, there is an overabundance of data on when, where, and what kind of ARs have occurred.  In the context of this letter we ask and answer the questions:  How many ARs of a given Hale class occur?  How does the magnetic class of ARs vary with solar cycle?

\section{Methods}
The reported AR data were obtained using the NOAA Active Region (NAR) database which is available as part of the Yohkoh database in the SolarSoft IDL software distribution.  The database contains information extracted from the daily Solar Region Summary (SRS) which are daily reports on solar activity jointly prepared by the U.S. Department of Commerce, NOAA, the Space Weather Prediction Center, and the USAF.  The multiple observatories in the USAF Solar Optical Observing Network (SOON) independently determine the location of sunspot groups, their areas, and magnetic configuration, and the combined reports are verified against additional sources, including other ground-based and space-based observatories such as the Solar and Heliospheric Observatory (SOHO) and the Solar Dynamics Observatory (SDO) (private communication with K.S. Balasubramaniam, 2016 February).  Each summary contains the heliographic latitude and longitude, Carrington cycle longitude, sunspot area corrected for projection, sunspot number, longitudinal extent, the Modified Zurich sunspot classification \citep{mcintosh90}, and the Mount Wilson magnetic classification of the ARs present on the Sun that day.  The USAF observing procedures are described in \citet{afwa13}.

The NAR database contained significant errors in the reported magnetic configurations prior to January 2016 from a bug in the code which generates the weekly database files.  These errors have been fixed and the NAR now contains the same information as reported in the SRS (private communication with S. Freeland and D. Zarro, 2015 December - 2016 January).

We consider the period between 1992 January 1 and 2015 December 31 which covers the end of solar cycle 22, all of solar cycle 23, and the majority of solar cycle 24, during which 5468 unique active region numbers were assigned.  NOAA active region numbers in the SRS are truncated to four digits, so this was accounted for by adding the remaining digits for all ARs occurring after 9999.  For each AR appearing in the database, the representative magnetic configuration and latitude were taken from the day the AR achieved its maximum sunspot area.  ARs present on the Sun for multiple rotations were not accounted for and they are present in these statistics as separate ARs.

\begin{deluxetable}{rrr}
	\tablecaption{Magnetic Configuration of Active Regions in Sample\label{tbl}}
	\tablehead{\colhead{Configuration} & \colhead{Number} &\colhead{\% of Total} }
	\startdata		
		\alp & 1064 & 19.46\\
		\alp\gam & 2 & 0.04\\
		\alp\delt & 0 & 0.00\\
		\alp\gam\delt & 1 & 0.02\\
		\hline
		\bet & 3512 & 64.23\\
		\bet\gam & 599 & 10.95\\
		\bet\delt & 46 & 0.84\\
		\bet\gam\delt & 239 & 4.37\\
		\hline
		\gam & 2 & 0.04\\
		\gam\delt & 2 & 0.04\\
		Unclassified & 1 & 0.02\\
		\hline
		Total \alp\ Classes & 1067 & 19.51\\
		Total \bet\ Classes & 4396 & 80.40\\
		\hline
		Total & 5468 & 
	\enddata
\end{deluxetable}

\begin{figure}
	\begin{center}
		\includegraphics[width=3.5in]{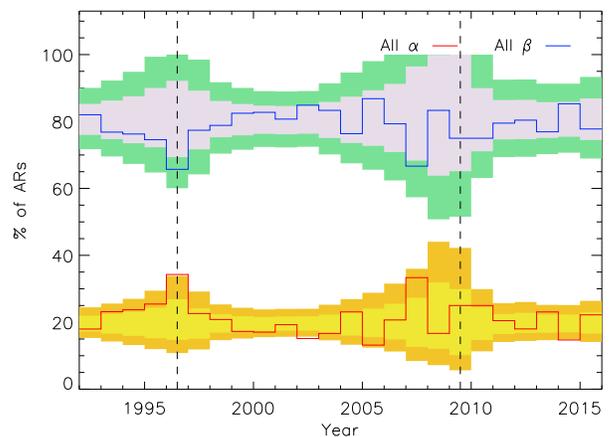}
	\end{center}
	\caption{The fraction of ARs in the \alp\ (red) and \bet\ (blue) classes each year.  The light and dark shaded regions indicate the 2- and 1-$\sigma$ Poissonian confidence interval for a model where \alp\ and \bet\ classes are produced in the constant fractions listed in Table \ref{tbl}.  The fractions are fairly constant in time, however the years 1996 and 2007 near solar minimum show an increase (decrease) in the fraction of \alp\ (\bet) classifications which are significant at the 1- to 2-$\sigma$ level.}
	\label{fig:ab_percents}
\end{figure}

\section{Results}
Table \ref{tbl} gives the number and proportion of magnetic configurations that occur in the sample.  The \bet\ classes account for the majority of ARs, in total 80.40\% of the sample.  The \alp\ classes account for 19.51\%, although almost all of these regions are categorized as simple \alp s.  All other AR classes (\gam, \gam\delt, and unclassified regions) account for a fraction of a percent of the sample.

\begin{figure}
	\begin{center}
		\includegraphics[width=3.5in]{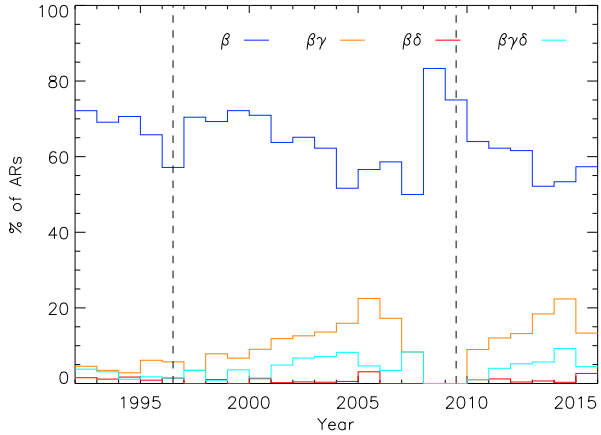}
	\end{center}
	\caption{Breakdown of the \bet\ classes of the sample for each year.  The \bet, \bet\gam, and \bet\gam\delt\ classifications show a large variation over the last two solar cycles.}
	\label{fig:b_percents}
\end{figure}

\begin{figure}
	\begin{center}
		\includegraphics[width=3.5in]{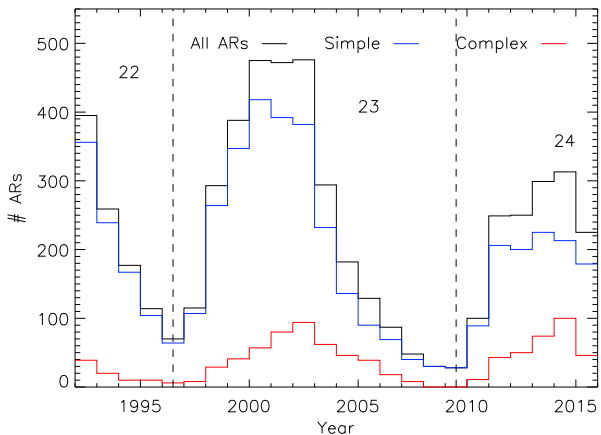}
	\end{center}
	\caption{Histograms of all ARs (black), ARs with simple configurations (blue), and ARs with complex configurations (red) binned by year.  For solar cycles 23 and 24, the highest number of complex configurations is reached the same year as the total number of ARs, but the peak in simple configurations occurs 1-2 years earlier.}
	\label{fig:AR_numbers}
\end{figure}

\begin{figure}
	\begin{center}
		\includegraphics[width=3.5in]{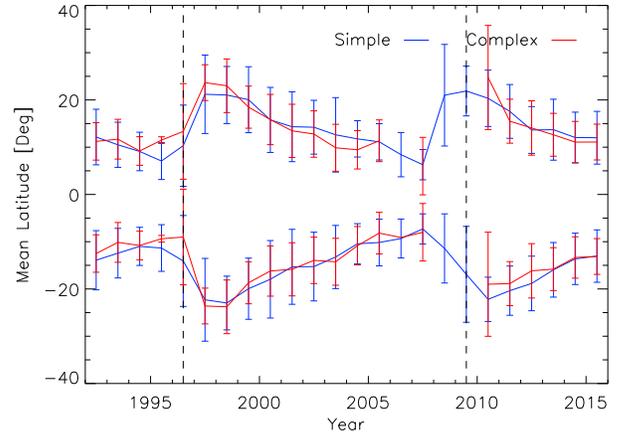}
	\end{center}
	\caption{Average latitude for ARs with a simple classification (blue), and with a complex classification (red) binned by year, determined separately for the northern and southern hemispheres.  Complex ARs do not appear at significantly different latitudes than simple ARs.}
	\label{fig:AR_latitude}
\end{figure}

In Figure \ref{fig:ab_percents} we show the variation of the two main AR classes, \alp\ and \bet, with time.  The sample statistics are binned by year to smooth over small variations and low statistics.  We plot the fraction of the \alp\ classes (red) and \bet\ classes (blue) relative to the total AR number.  The vertical dashed lines in this and each following plot mark the year the lowest sunspot number was reach between each cycle.  The fraction of of all \alp\ classes and \bet\ classes is roughly constant in time at a 20\%/80\% proportion respectively.  The shaded regions around each line graph indicate the 1- and 2-$\sigma$ Poissonian confidence intervals.  To determine the confidence interval and its mean we assume a model where the AR classes occur independently of one another in the constant proportions listed in Table \ref{tbl}.  The interval is calculated using the approximate method of \citet{gehrels86}.  An increase in the observed fraction of \alp\ classes, and a decrease in the \bet\ classes, at the 1- to 2-$\sigma$ level is noticeable for the years 1996 and 2007 near the minima in activity.

Figure \ref{fig:b_percents} shows the breakdown of the \bet\ sub-classes as a fraction of the total number of ARs for each year.  There is a large variation in the relative fractions of individual \bet\ classes produced as a function of time.  At the peak of solar cycles 23 and 24, \bet\gam\ and \bet\gam\delt\ configurations account for more than 30\% of the ARs classifications.  During the lowest activity levels reached between solar cycle 22 and 23 in 1996 about 10\% of ARs had complex configurations, and during the very quiet minimum between solar cycles 23 and 24 in 2009 there were no ARs with complex configurations.  While this time period does not contain the maximum of solar cycle 22, we might expect to see an increased level of complex configurations during the slow declining phase of the cycle during the years 1992 to 1996, but the fraction of complex regions reported was almost as low as the minimum.

We only consider ARs with complex configurations (\bet\gam, \bet\delt, \bet\gam\delt, \gam, \gam\delt) and those with simple configurations (\alp, \bet) for the remainder of this section.  In Figure \ref{fig:AR_numbers} we show the number of ARs categorized as simple (blue) and complex (red), along with the total number of ARs (black), for each year.  The cycles are demarked by the vertical dashed lines.  Considering the activity cycle for simple and complex regions separately, we can see that from both solar cycle 23 and 24, the peak number of simple ARs occurs a year or two earlier in the cycle than the peak for all ARs.  The peak number of complex ARs is reached the same year as the peak number of all ARs, and for solar cycle 23 the numbers of complex ARs do not decrease as quickly, leading to the higher relative fraction late in the cycle seen in Figure \ref{fig:b_percents}.

In Figure \ref{fig:AR_latitude} we show the average latitude for simple and complex classifications that occurred in the north and south hemispheres, where the error bars show the standard deviation of the AR latitudes in each of the yearly bins.  The distribution of AR types with latitude is similar for simple and complex ARs.

\section{Discussion}
The subjective nature of AR classification makes it difficult to distinguish real changes in the complexity of ARs from changes in the way observers have classified regions.  The \alp\ and \bet\ classes assigned by \citet{hale38} and in the SRS shown in this work both show fairly flat trends with solar cycle.  However \citet{hale38} classified 38\% of regions as \alp\ and 62\% as \bet, while in this work we find that on average about 20\% are \alp-type and 80\% are \bet-type, though for a brief periods near the solar minima we see a similar distribution.  \citet{hale38} assigned a much smaller fraction of complex classifications, only 5\%, while in this work complex classifications sometimes account for over 30\% of all ARs.

Changes in instrumentation, observation quality, and observer bias may be responsible for the overall shift in the fractions of \alp\ and \bet\ classifications and the increase in the fraction of complex ARs identified.  Early observers at Mount Wilson relied on visual interpretation of polarized Zeeman profiles from only the small regions covered by the slit.  With the introduction of scanning magnetographs it became possible to map the full disk of the Sun to give a more complete view of the magnetic morphology.  The bipolar nature of ARs is clearly shown in magnetograms even when sunspots are not visible, which may result in fewer instances of the \alp\ class assigned in modern observations.  Further improvements in the spatial resolution of magnetograms may be responsible for the increase in the number of complex ARs identified.  The SOHO, launched in 1996, and SDO, launched in 2010, have revolutionized the observation of AR magnetic fields by providing high resolution magnetic field maps at high cadence, making it easier to identify regions of mixed magnetic polarity and \delt-sunspots.  The addition of data from these two facilities may be responsible for what appears to be a gradual increase in the fraction of complex ARs identified from 1992 to 2016.  SDO fully resolves AR scale magnetic flux, so further increases in spatial resolution are unlikely to change AR classification significantly.

Each individual class in the modern classification shown in Figure \ref{fig:b_percents} shows variation during the two most recent solar cycles which is at odds with the results in \citet{hale38} which show very little variation in any of the classifications as a function of time.  The variation in the assignment of complex classifications cannot be attributed only to observer bias because similar behavior is seen during both solar cycles 23 and 24.

The transition from simple \alp\ and \bet\ classes at the beginning of the cycle, to more complex \bet\gam\ and \bet\gam\delt\ classes at the peak and during the declining phase of the cycle may provide crucial evidence for how ARs form.  \citet{zirin87} presented three formation scenarios for \delt-regions based on the histories of a small sample: (1) emergence of a single AR complex containing a \delt-spot, (2) emergence of new flux inside an existing AR, and (3) emergence of separate ARs that collide as the result of evolutionary forces.  All of three scenarios can be produced by the collision of magnetic flux systems and the timing of the collision determines which scenario is seen.  Scenario (1) can also be produced by the emergence of a monolithic system of contorted magnetic flux like the \bet\delt-like AR flux emergence modeled in \citet{fang15}.

The solar cycle variation in AR magnetic complexity supports the idea that complex ARs are produced by the collision of magnetic flux at the photosphere due to the higher frequency of magnetic flux emergence.  Both simple and complex regions have the same latitudinal dependence on solar cycle, indicating that they arise from the same reservoir of flux in the solar interior.  One might expect complex, monolithic magnetic structures to occur uniformly throughout the solar cycle, but if the majority of complex ARs are produced by the collision of separate systems of emerging flux, more complex ARs should occur during the declining phase of the cycle when the latitudinal width of the activity band decreases \citep{hathaway}.  A larger fraction of complex classes should also be seen in stronger cycles where more flux emergence is occurring, but the bias introduced by increasing spatial resolution makes this difficult to support.

\section{Conclusions}
For each unique AR reported between 1992 and 2015, we have taken the Mount Wilson classification for the day they reached their peak area.  We find that \alp (\bet) -type ARs make up 20\% (80\%) of ARs except for a brief time during the solar minima when they make up 35\% (65\%).  We show that there is variation in AR complexity as a function of solar cycle as indicated by the classifications assigned during this time, contrary to the flat trends with solar cycle shown in \citet{hale38}.   A larger fraction of simple regions occur early in the solar cycle, while at solar maximum and during the declining phase over 30\% of regions were given complex classifications.  Complex ARs and simple ARs show no significant difference in distribution with latitude during the solar cycle, implying that they originate from the same reservoir of flux in the solar interior.  This trend in complexity with solar cycle has a clear implication for the origin of complex ARs: they are produced by the collision of different magnetic systems when the frequency of flux emergence is high, rather than the emergence of complex, monolithic magnetic structures.

\acknowledgments
This research was supported by an appointment to the NASA Postdoctoral Program at Goddard Space Flight Center administered by the Universities Space Research Association through a contract with NASA.

\end{document}